\documentclass[fleqn,12pt,twoside]{article}
\usepackage{espcrc1}

\readRCS
$Id: espcrc1.tex,v 1.2 2004/02/24 11:22:11 spepping Exp $
\usepackage{graphicx}
\usepackage[figuresright]{rotating}
\usepackage{amssymb}

\newcommand{\AmS}{{\protect\the\textfont2
  A\kern-.1667em\lower.5ex\hbox{M}\kern-.125emS}}

\begin{document}

\title{\bf Very high energy observations of the BL Lac objects \\ 3C 66A and OJ 287 \\}

\author{T. Lindner
\address
[McGill]{Department of Physics, McGill University \\Montreal, QC H3A 2T8, Canada}\thanks{present address: Department of Physics and Astronomy, University of British Columbia, Vancouver BC V6T 1Z1, Canada}, 
D.S. Hanna \addressmark[McGill], 
J. Kildea\addressmark[McGill]\thanks{present address: Fred Lawrence Whipple Observatory, Harvard-Smithsonian Center for Astrophysics, Amado AZ, 85645, USA}
,
J. Ball 
\address
[UCLA]{Department of Physics and Astronomy, University of California, Los Angeles \\ Los Angeles, CA 90095, USA}\thanks{present address: Gemini North, Hilo, HI 96720, USA},
D.A. Bramel
\address
[Barnard]{Department of Physics and Astronomy, Barnard College, Columbia University\\ New York, NY 10027, USA}\thanks{present address: Interactive Brokers, Greenwich, CT 06830, USA}
,
J. Carson\addressmark[UCLA]\thanks{present address: Stanford Linear Accelerator Center, Menlo Park, CA 94025, USA}
, 
C.E. Covault 
\address
[Case]{Department of Physics, Case Western Reserve University\\ 
Cleveland, OH 44106, USA},
D. Driscoll\addressmark[Case],
P. Fortin\addressmark[Barnard],
D.M. Gingrich
\address
[Alberta]
{Centre for Particle Physics, Department of Physics, University of Alberta\\
Edmonton, AB T6G 2G7 and TRIUMF, Vancouver, BC V6T 2A3, Canada}
A. Jarvis \addressmark[UCLA],
C. Mueller \addressmark[McGill],
R. Mukherjee \addressmark[Barnard],
R.A. Ong \addressmark[UCLA], 
K. Ragan \addressmark[McGill],
R.A. Scalzo 
\address 
[Chicago]
{Department of Physics, University of Chicago, Chicago, Il 60637, USA}
\thanks{present address: Lawrence Berkeley National Laboratory,
Berkeley, CA 94720 USA}
,
D.A. Williams
\address
[UCSC]
{Santa Cruz Institute for Particle Physics, University of California, Santa Cruz\\ Santa Cruz, CA 95064, USA},
J. Zweerink \addressmark[UCLA]
}

\maketitle

\newpage 

\begin{abstract}

\noindent
{\bf Abstract}

Using the Solar Tower Atmospheric Cherenkov Effect Experiment (STACEE), we have
observed the BL Lac objects 3C 66A and OJ 287. These are members of the class
of low-frequency-peaked BL Lac objects (LBLs) and are two of the three LBLs 
predicted by Costamante and Ghisellini \cite{Costamante} 
to be potential sources of very 
high energy ($>$ 100 GeV) gamma-ray emission. The third candidate, BL Lacertae,
has recently been detected by the MAGIC collaboration \cite{albert07}.
Our observations have not produced detections; we calculate a 99\% CL upper 
limit of flux from 3C 66A of 0.15 Crab flux units and from OJ 287
our limit is 0.52 Crab. These limits assume a Crab-like energy spectrum with 
an effective energy threshold of 185 GeV.

\end{abstract}

\section{Introduction}

The field of very high energy (VHE) gamma-ray astronomy is a relatively 
young discipline that is concerned with the study of astrophysical sources 
of gamma rays with energies above 100 GeV.
At these energies, all gamma-ray detections are indirect; they are made 
using ground-based telescopes which measure components of 
the air showers caused by the gamma rays.
The lowest energy thresholds are achieved by telescopes which measure 
the Cherenkov
light produced by particles in the showers. 
Detectors which record the arrival of the air shower particles themselves 
achieve wider fields of view and larger duty factors but operate at higher
thresholds.
The first reliable detection of an astrophysical source using the atmospheric
Cherenkov technique
was that of the 
Crab Nebula, made in the late 1980's, by the Whipple collaboration 
\cite{weekes89}.
Since then there has been rapid progress in the field (for a recent review 
see \cite{ong05}).

The first extra-galactic VHE 
source to be detected was the blazar Markarian 421 
\cite{punch92}. 
Since its detection in 1992, more than a dozen other blazars have been 
detected at TeV energies  and they constitute almost all of 
the known extra-galactic VHE sources.
At lower (GeV) energies, 66 of the 271 sources in the EGRET catalog 
\cite{hartman99} have been identified as belonging to the blazar class, 
again constituting the majority of the identified extra-galactic sources.

Blazars are members of a class of Active Galactic Nuclei (AGNs). 
Simply described, an AGN is believed to comprise a super-massive 
($10^7 - 10^9~M_{\bigodot}$)
black hole surrounded by an accretion disk at the centre of a host galaxy.
Relativistic jets emerge along the spin axis of the AGN.
Blazars are those AGNs which have one of their jets pointed towards the Earth.

In the leading blazar paradigm, VHE gamma rays are produced by inverse-Compton
(IC) scattering of low energy photons by a population of high energy electrons
which have been shock-accelerated in the jet. 
These electrons, moving in the local magnetic fields, 
also produce synchrotron radiation. 
This scenario
leads naturally to a double-hump spectral energy distribution (SED) with a 
low energy synchrotron peak and a higher energy IC peak.
Blazars are often classified by the location of the synchrotron peak.
Low-frequency-peaked blazars (LBLs) have this peak in the radio or 
optical band while
for high-frequency-peaked blazars (HBLs), it is in the X-ray band.
Costamante and Ghisellini \cite{Costamante} have studied a large number of blazars 
with the aim of predicting which ones could be detectable at TeV energies.
All of the blazars detected by VHE telescopes, before and after  
publication of their study, have satisfied their search criteria.
Until very recently \cite{albert07}, all were members of the HBL class. 

There are three members of the LBL class which are included in their list 
of candidate TeV emitters: 3C 66A, OJ 287 and BL Lacertae.
All three have lower X-ray fluxes than the known TeV sources but they have 
relatively large radio fluxes. 
It is the combination of high energy electrons (implied by large X-ray fluxes)
and large numbers of seed photons (which make up the large radio flux) which 
can give rise to a significant TeV gamma-ray output.  
Thus it is possible that the large radio flux can compensate for
the relatively low X-ray flux, with the result that a detectable flux of TeV 
photons is produced.

At the time of the observations reported here, none of the three LBL 
candidates had been reliably detected in the VHE band.
Recently MAGIC \cite{albert07} has reported a detection of BL Lacertae
based on 
22.2 hours of data acquired in 2005 and a non-detection of the source 
based on 26.0 hours of data acquired in 2006. 
The detected flux (about 3\% of the Crab above 200 GeV) is evidence
in favour of the arguments of the previous paragraph.
The non-detection in 2006 is a reminder that blazars are time-variable and 
it is not possible to guarantee that any given observation will result in 
a detection. 

In the case of 3C 66A and OJ 287,
the lack of TeV detection could be due 
to the large distances to the objects. 
The redshift for 3C 66A is 0.444 (although this is not well established - see
\cite {bramel}) and for OJ 287 it is 0.306, and it is possible that the 
gamma-ray fluxes are
attenuated by pair-production with the intervening extra-galactic background 
light (EBL) \cite{stecker92}.

We have attempted to detect these two LBLs using the 
Solar Tower Atmospheric Cherenkov Effect Experiment (STACEE) 
detector which operates at 
a lower energy threshold than the earlier generation of atmospheric 
Cherenkov telescopes.
Given the steeply falling spectrum of known TeV blazars and/or the significant
energy dependence of the EBL absorption effect, a detector
with a lower energy threshold ($\sim$ 100 GeV) 
would be better suited to detect these sources,
should they be VHE emitters.

\section{The STACEE Project}

\begin{figure}[h]
\centerline{\includegraphics[width=1.0\textwidth]{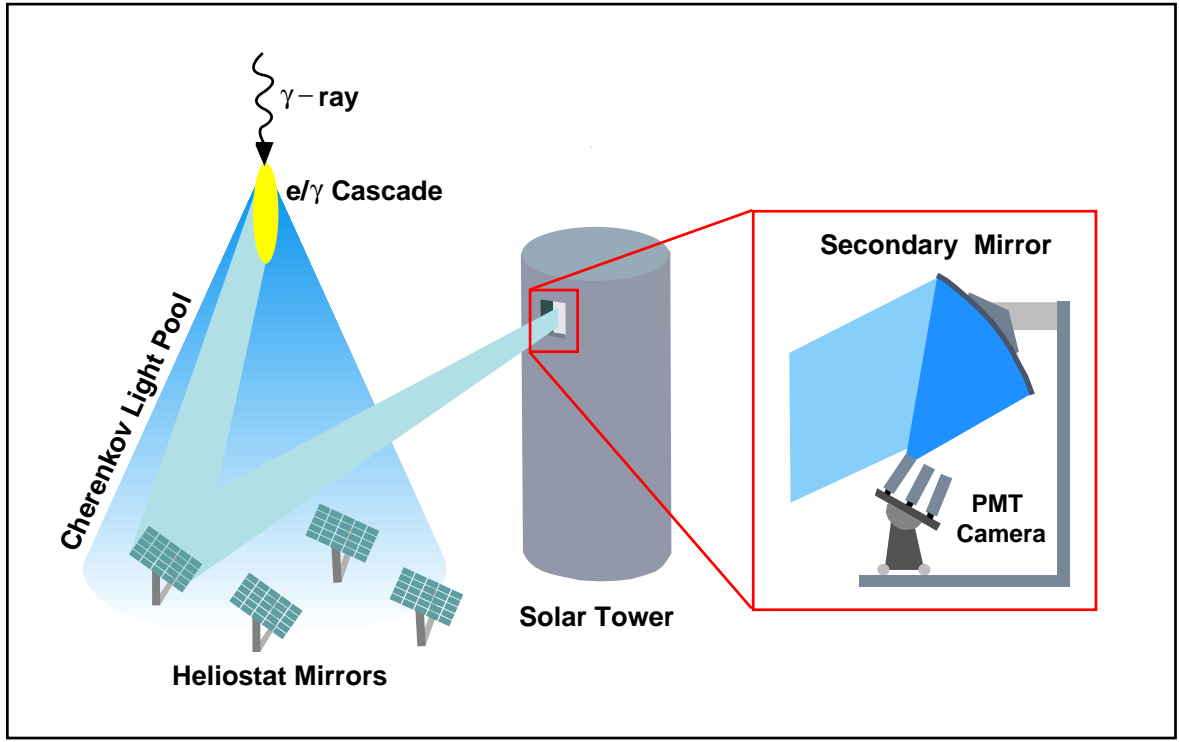}}
\caption{\label{concept} Concept of the central tower gamma-ray detector.
Cherenkov light from the air shower initiated by the incident
gamma ray is directed by large heliostats towards secondary optics located
on the tower. These optics focus the light from each heliostat onto a 
corresponding photomultiplier tube.}
\end{figure}

The STACEE detector
is installed at the National Solar 
Thermal Test Facility (NSTTF) at Sandia National 
Laboratories in Albuquerque, New Mexico (34.96 N, 105.51 W).
Like most other VHE gamma-ray detectors, 
it uses the atmospheric Cherenkov technique
to detect astrophysical gamma rays, but, unlike most Cherenkov telescopes,
it is not an imaging detector.
STACEE belongs to a class of wavefront sampling detectors which use the large
steerable mirrors (heliostats) of solar power research facilities to reflect
Cherenkov light onto secondary mirrors located on a central tower. 
The secondary mirrors focus the light onto photomultiplier tubes (PMTs),
one per heliostat.
The concept is illustrated schematically in figure~\ref{concept}. 
Other such detectors were operated in France \cite{celeste}, Spain 
\cite{graal} and the US \cite{cactus}.
A more complete discussion of these solar heliostat telescopes
can be found in \cite{smith}.

The STACEE detector has been described previously 
\cite{chantell,hanna,gingrich}; 
we give here a brief description of its configuration relevant to the 
data presented here.

\subsection{Heliostats and Secondary Optics}

\begin{figure}[p]
\centerline{\includegraphics[width=1.0\textwidth]{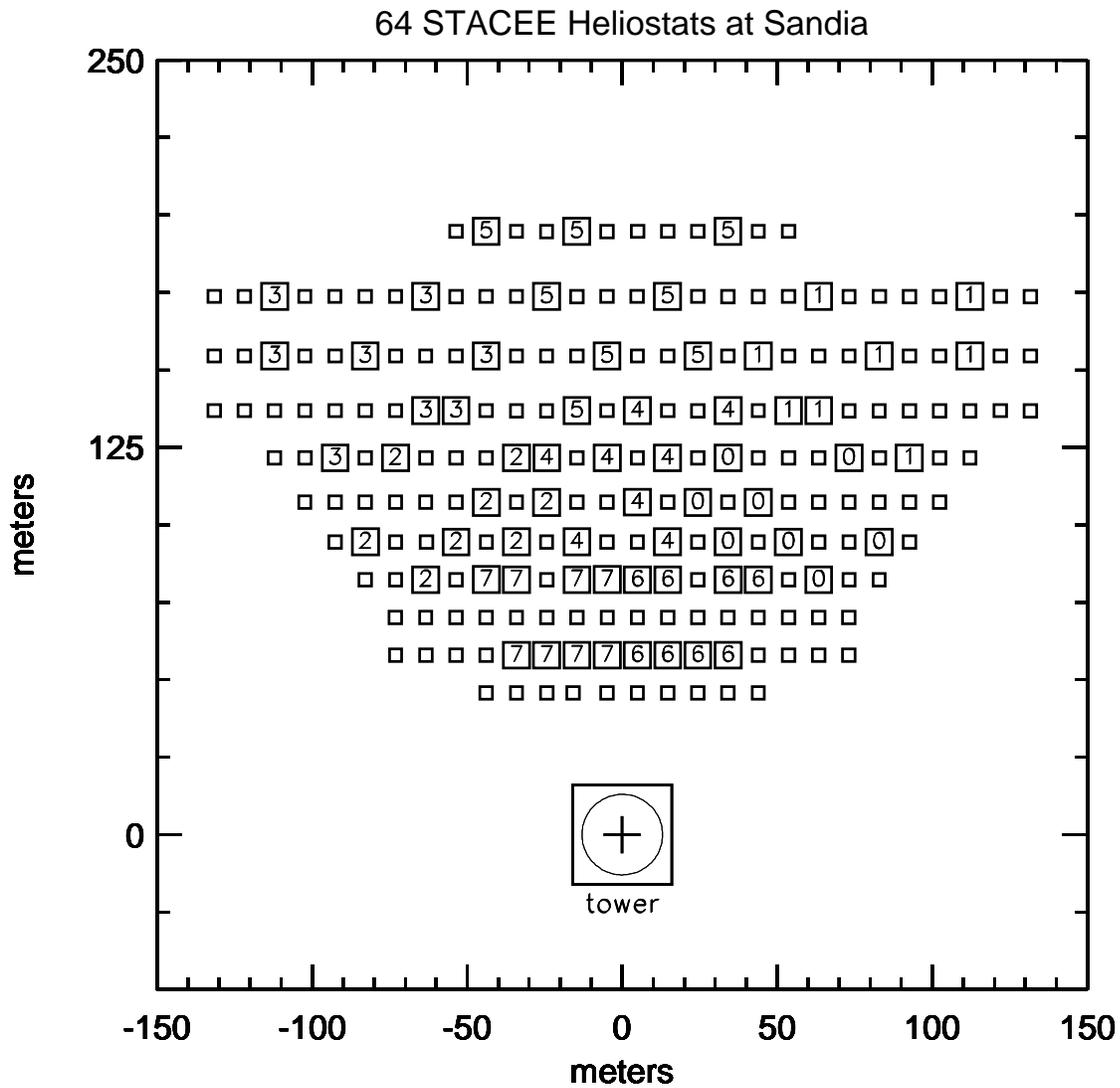}}
\caption{Map of the STACEE heliostat field.
Each square denotes a heliostat and the ones used by STACEE are numbered
from 0 through 7, according to the trigger cluster to  
which they are assigned.
The orientation of the field is such that the heliostat rows are in an 
east-west direction and the tower is on the southern edge.
\label{hfield}}
\end{figure}

There are 212 heliostats in the NSTTF array, 
each with a mirror area of $37~m^2$.
STACEE uses 64 heliostats distributed throughout the field, 
as shown in figure~\ref{hfield},
and grouped into eight clusters of eight heliostats each.
Light from the heliostats is directed towards a central tower where five 
secondary mirrors and associated cameras are located.
Three cameras, each with 16 channels, 
are located at the 160-foot level of the 
tower and two more, each with eight channels, are located at the 120-foot level.
Each camera is at the focal point of a spherical f-1.0 secondary mirror (2.0 
m diameter for the 16-channel cameras and 1.1 m diameter 
for the 8-channel cameras) which collects 
and focusses Cherenkov light reflected 
from the heliostats onto PMTs in the camera.
Light concentrators coupled to the PMTs increase their effective areas and 
define their fields of view.

\subsection{Electronics and Trigger}

Pulses from the PMTs are first sent to high-speed
amplifiers where the signal size is increased. 
Each amplifier channel produces two outputs; one is sent to a discriminator,
followed by digital trigger logic and the other is digitized by an 8-bit 
flash analog-to-digital converter (FADC) (Acqiris DC-270)
operating at 1 GS/s.
The front-end amplifiers are 
commercially available NIM modules (Phillips 776). Two units, each with a 
gain of 10, are cascaded for each PMT. 

The trigger is a custom-made device which uses field-programmable-gate-arrays
(FPGAs) to delay each pulse (to account for the relative geometry of the 
Cherenkov wavefront and the heliostats) 
and require a majority coincidence of PMT signals
to form a first-level trigger.
The delays are dynamic; because of the earth's rotation, a given source 
appears to 
move across the sky, so the relative timing of each channel needs to be 
continuously modified to maintain a tight trigger coincidence window.
STACEE uses a window of 12 ns effective width. 
To keep the inter-channel delays reasonably short, the trigger operates in 
two stages. 
For the first stage, the 64 channels are grouped into eight groups of eight channels, 
corresponding to heliostats that are close together on the field 
(see figure~\ref{hfield}).
A given number of these channels (typically five) in a group 
is required to be above threshold for that group to trigger.
For the second stage a given number of groups (also typically five)
is required to fire before reading out the detector. 
This two-stage requirement has the effect of selecting light pools that
are uniform
over a large part of the heliostat field.
Gamma-ray showers, which  result in smooth light pools,
are more likely to satisfy such a selection criterion than
hadron-initiated showers, which tend to be clumpier.

\section{Data Taking and Analysis}

\subsection{Observing Strategy}

STACEE employs an \lq ON-OFF\rq~ observing strategy. 
`ON' runs of 28 minutes wherein the source is 
tracked at the centre of the field
of view are alternated with `OFF'
runs where a patch of sky at identical declination 
but 30 minutes ahead or behind the source in right ascension is observed. 
The basic data unit is, then, a pair of such runs. 
The idea is that many backgrounds (for example terrestrial light sources)
depend on local coordinates, and their effects will be the same in both runs. 
Additionally, and most importantly, the rate of background showers from 
charged cosmic rays will be the same for both runs, so that one can infer 
the gamma-ray flux from the difference in count rates between the two.

Each night, prior to taking data on a given source, special runs, where 
the trigger rate is measured as a function of increasing 
threshold setting, are performed.
The trigger threshold is set just 
above the point where noise triggers, resulting from random
coincidences of night sky background photons, cease to dominate the triggers 
from air showers. 
Typically we run at a threshold of 5-6 photoelectrons per channel.
This number depends on atmospheric conditions, which affect the amount of
scattered light.

\subsection{Data Quality Cuts}

Data analysis proceeds in stages. 
As part of the first stage, runs where weather conditions were poor
are rejected, as are those where log files reveal periods of
unstable tracking by one or more heliostats.
Initial offline analysis also involves comparing the two runs of a pair. 
Count rates and currents for each channel and trigger group are required
to be consistent between the two runs.
Portions of runs where such quantities are not consistent
are eliminated from further analysis.
This criterion is applied to both runs of a pair; 
for example, if a cluster trigger 
rate is essentially constant for one run but deviates for a short time 
during the other run of the pair ($e.g.$ due to the passage of an airplane or
a small cloud), the offending interval is removed from both runs.

\subsection{Field Brightness Correction}

A very important analysis task is that of correcting for the relative 
brightness of the ON and OFF fields.
The effect is well-known in ground-based gamma-ray astronomy and different
methods have been used by different groups to cope with it
\cite{cawley}.

The ON-OFF observing strategy assumes that the difference between the total 
number of events from the ON run and that from the OFF run is
due to a flux of gamma rays from the targeted source.
Any background suppression and other analysis techniques are primarily used to 
improve the statistical significance of the gamma-ray excess.
However, if the night-sky background (NSB) is different for the two observing fields,
$e.g.$ due to one or more bright stars in one of the fields, a difference in 
count rates can arise from promotion effects.
Promotion occurs when an air shower, having deposited enough 
Cherenkov light in the various channels of the detector to be just below 
threshold, is raised over threshold by the addition of one or more NSB photons
which arrive during the trigger window.
If the extra brightness is in the ON field, a spurious signal can develop or, 
if it is in the OFF field, a genuine signal can be lost or weakened.
With STACEE data we correct for this effect using the FADC traces. 
In the following discussion we will assume that the ON field has the
extra NSB.

An obvious solution to correct for a brighter ON-source field 
would be to add extra photo-electron signals, 
corresponding to the ON-OFF difference in photo-currents, to the 
OFF-field traces for each channel.
This technique is called \lq software padding\rq.
One could do this with simulated single photo-electron waveforms,
but a detailed understanding of the pulse shape and its 
fluctuations at low charge levels is required.
Instead, we adopt an empirical approach. 
It is an observational fact that, because of the AC coupling at the front end
of the FADC and the high rates of NSB photons, which result in pile-up effects,
the baseline FADC trace looks very much like random noise and can be well 
described by a single parameter, its variance.
It is also true that the variance is linearly related to the photo-current.
To equalize the ON and OFF FADC traces in a given channel, we 
compute the difference in baseline variance between ON and OFF and add an 
additional FADC trace having this variance to the OFF trace.
By construction, the ON and OFF FADC baseline traces now have equal variances 
but any coherent air shower signal has not been affected.

\begin{figure}
\centerline{\includegraphics[width=0.8\textwidth]{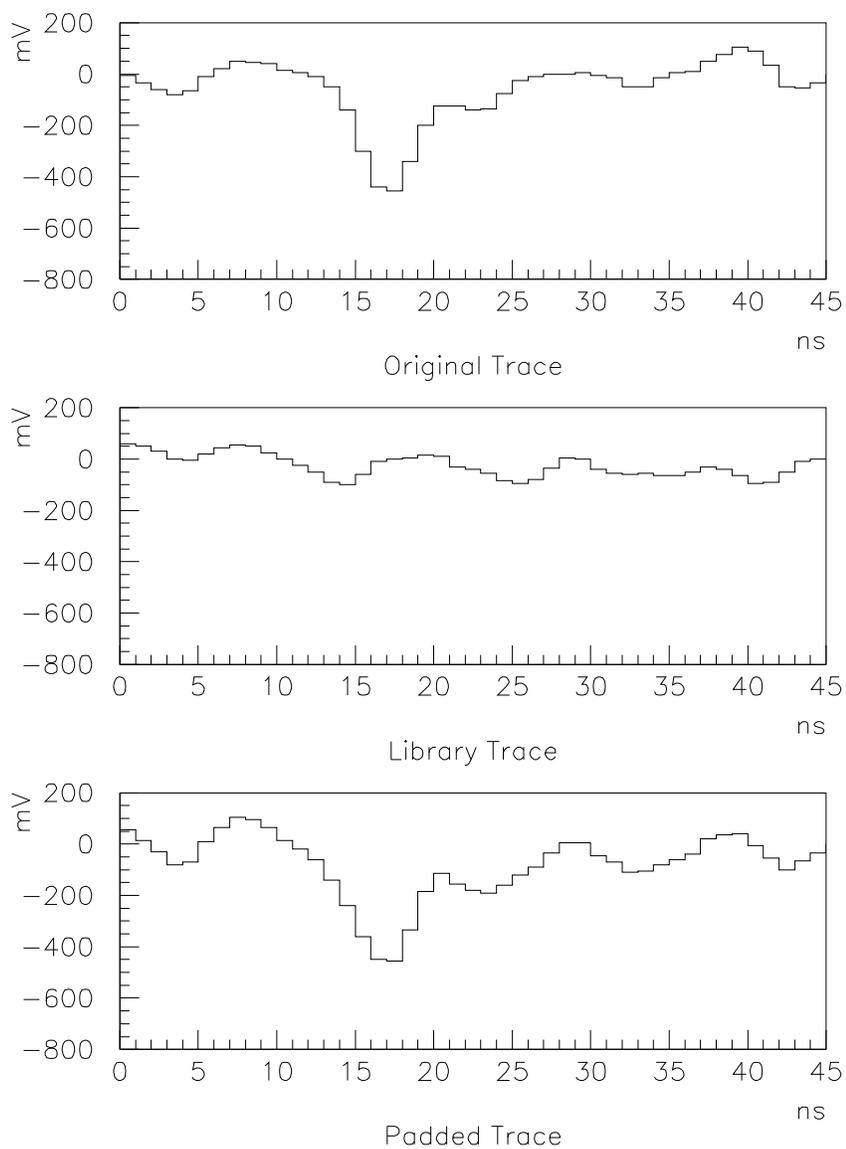}}
\caption{Schematic illustration of the `library padding' concept. 
An FADC trace with appropriate baseline variance, taken from a library of such traces, is added to the trace from
a given event and channel to produce a pulse with larger baseline 
variance.
All variances are calculated from the portion of the FADC trace which 
precedes the triggered pulse.}
\label{padding}
\end{figure}

The added trace is taken from a large library of such traces. 
The library traces are made by illuminating a PMT with a variable
intensity light-emitting diode and triggering the read-out with a pulse 
generator.
Thus the traces contain only controlled levels of baseline fluctuations.
This technique is called \lq library padding\rq and was first used in 
\cite{scalzo}; the concept is illustrated 
schematically in figure~\ref{padding}.

The padding process is followed by a re-imposition of the trigger in software. 
This must be done using a higher threshold than was used during data
collection because, even though the OFF field has now been artificially 
brightened, any sub-threshold showers that would have triggered the 
experiment had the
brightness been there in the first place cannot be recovered; they are 
not in the data set. 
Thus one must remove the corresponding showers from the ON-field data.
The increment to the trigger threshold has been empirically determined by 
examining a series of ON-OFF pairs taken with bright stars of different 
magnitudes.
   
\subsection{Hadron Rejection}

After the data quality criteria are applied,
we are left with data sets made up of series 
of triggered events, each containing 64 FADC traces, one
per heliostat.
Each trace is 192 samples long and contains the digitized primary PMT pulse
near its centre. 
The times and charges of the pulses are extracted from these traces to be used
in later analysis steps and the traces themselves are used in the 
hadron rejection process.

One of the most important steps in the analysis
is that of rejecting showers 
caused by charged
cosmic rays, henceforth called hadronic showers.
We are working with a set of air showers where the ratio of 
gamma-induced air showers to hadronic showers is very small, the exact ratio 
depending on the gamma-ray source being observed.
As stated earlier, the flux of gamma rays can be estimated from the ON-OFF 
count rate difference, but it is important 
to improve on the gamma/hadron ratio
to enhance the statistical quality of this estimate.

STACEE is not an imaging detector, so the powerful hadron rejection techniques 
devised for such detectors \cite{hillas} cannot be applied to our data.
Instead, we use a scheme \cite{smith06} referred to here as 
the {\it grid alignment technique}, for reasons that will 
become clear. This method was adapted for use in STACEE 
\cite{kildea_a,lindner},
and has been tested using observations of the Crab Nebula 
\cite{lindner,kildea_b}  
where it has been shown to improve the signal from a 4.8 
$\sigma$ excess to 8.1 $\sigma$ in our 2002-2004 data set.
Stated differently, our sensitivity to the Crab is 1.62 $\sigma/\sqrt{hour}$.
This is to be contrasted to the sensitivity obtained during our first 
observation of the Crab \cite{oser} with the STACEE-32 detector \cite{hanna}
which was 1.03 $\sigma/\sqrt{hour}$.

Gamma/hadron separation using the grid alignment technique relies on the 
difference in shapes between 
the wavefronts of gamma-induced showers and hadronic 
showers at the energies of interest to STACEE. 
By \lq wavefront\rq, we mean the distribution 
in space and time of the Cherenkov 
photons which arrive at the detector.
Simulations show that gamma-induced showers have a smooth wavefront
with a shape that forms part of a sphere, the origin of which is at 
the position of shower maximum, 
the point at which the population of particles in the 
air shower reaches its maximum value.
Shower maximum 
is approximately 10-12 km above the detector for vertically incident
gamma-ray showers. 
By contrast, hadronic showers have much more sub-structure and their wavefronts
are not usually spherical. 
Calculating and selecting on the sphericity of showers is expected to be a 
useful tool in rejecting hadronic showers.

\begin{figure}
\centerline{\includegraphics[width=1.0\textwidth]{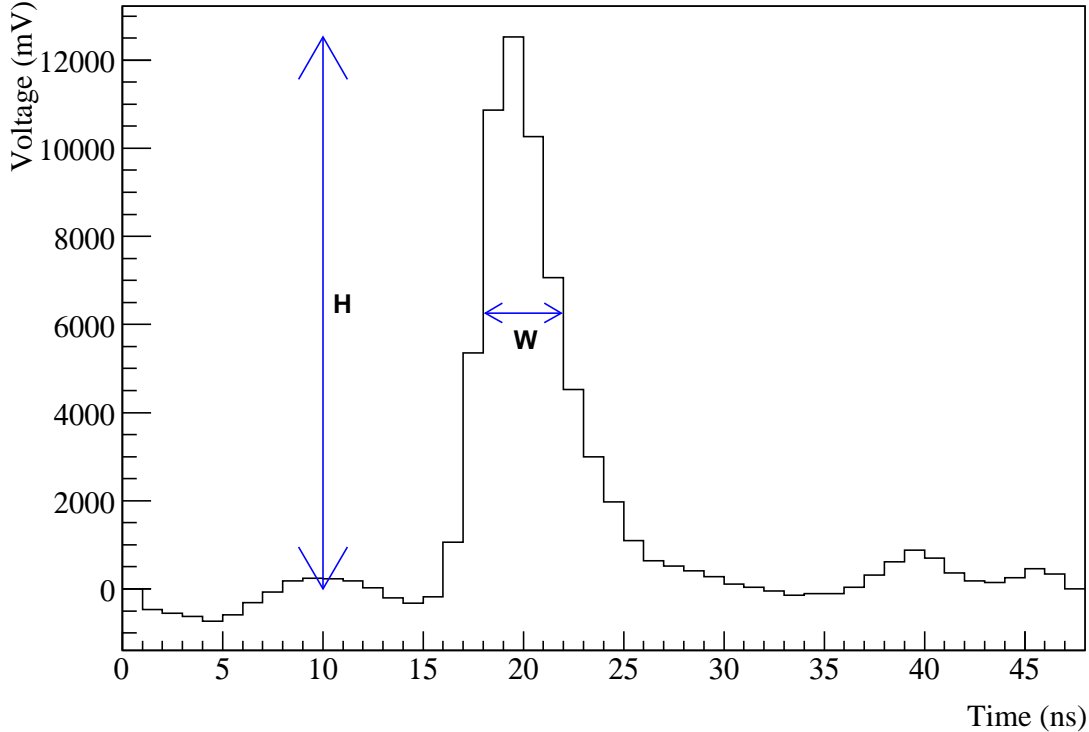}}
\caption{Sum of all FADC traces for a single, simulated, gamma-ray event.
The quantities, $H$ and $W$, used in the grid ratio method of hadron rejection 
are shown by way of illustration.
This plot was made using the position of the shower core that gives the largest
value of $H/W$.
Note that, before summing, all traces are pedestal-subtracted and inverted.\label{HW}}
\end{figure}

\begin{figure}
\centerline{\includegraphics[width=1.0\textwidth]{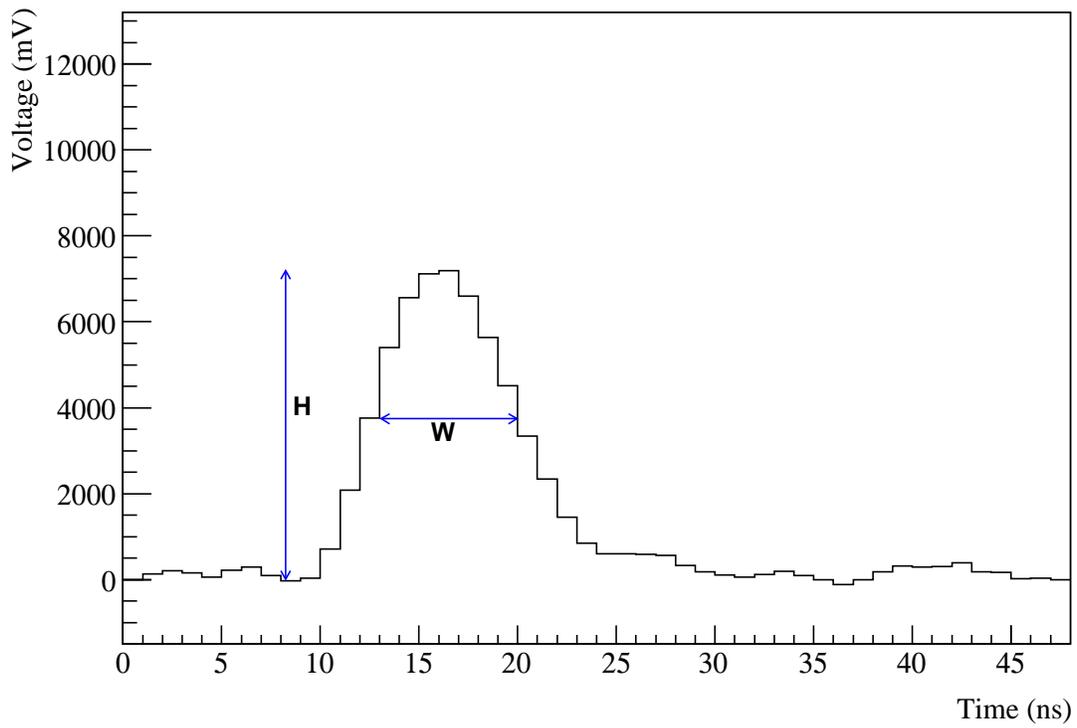}}
\caption{As in the previous figure but made with a shower core position 200 
metres from the optimal position used in the previous figure.\label{HW2}}
\end{figure}

To calculate sphericity it is necessary to know where the core of the shower
landed in the heliostat field.
This is the point at which the incident gamma ray would have impacted the 
field had it not interacted in the atmosphere and initiated the shower.
The core position is {\it a priori} unknown; for purposes of trigger timing,
it is assumed to be at the centre of the field, but its true value can 
be quite far away since timing tolerances in the STACEE 
trigger allow for a range of values.
To estimate more precisely the core position, we step through a grid of 
possible locations.
At each point on the grid we calculate the time of arrival of the wavefront 
at each heliostat
based on the geometry given by the core position and the location of shower 
maximum, 
assuming that it lies on the line connecting the core position and the 
targeted source.
The position of shower maximum 
along the line is adjusted for atmospheric depth 
according to the elevation angle of the source.
The differences of the expected 
arrival times from their observed values are used to 
shift the start times of each FADC trace and the traces are then summed.
For the correct core position, pulses from the different heliostats will add
coherently and produce a summed pulse with large amplitude and narrow width.
(See figure~\ref{HW}.)
For an incorrect core position, the amplitude will be smaller and the 
width will be larger, as shown in figure~\ref{HW2}.
The ratio of amplitude (height $H$) to width ($W$) of the summed pulses is used
as a figure of merit in the following steps and it is assumed that the 
hypothesized core position which maximizes $H/W$ is the best estimate of the 
true core position.

\begin{figure}
\centerline{\includegraphics[width=1.0\textwidth]{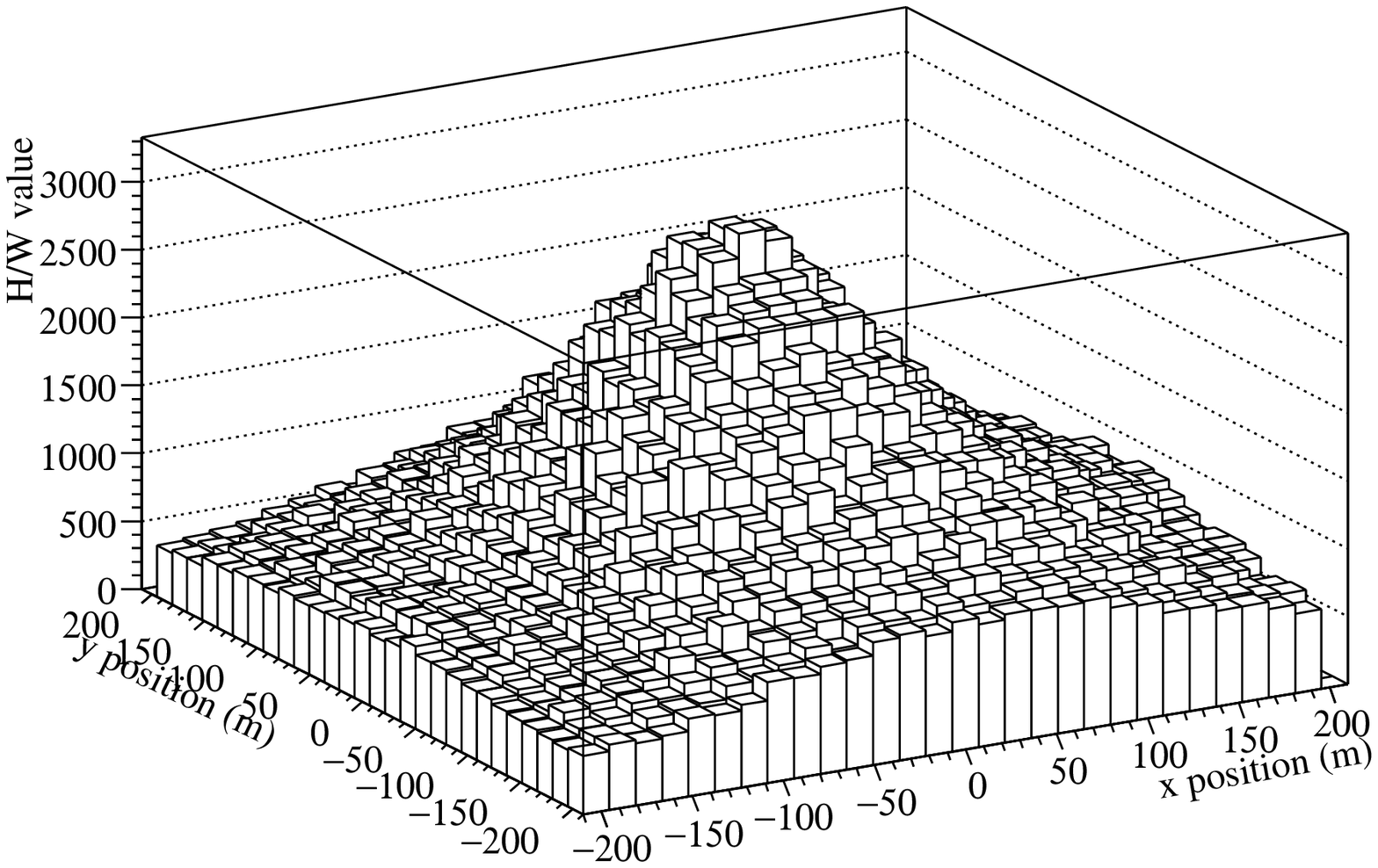}}
\caption{Illustration of the behaviour of the height-to-width ratio for the
sum of FADC traces as a function of hypothesised shower core locations.
This plot corresponds to a single simulated gamma-ray event and shows the 
characteristic peaking and symmetric behaviour expected of an electromagnetic
shower with a spherical wavefront.\label{HWg}}
\end{figure}

To find the core position, a 30x30 grid with 15 m pitch is stepped through 
and the location with the maximum value of $H/W$, $H/W_{max}$, is saved.
For gamma-ray showers one expects a peaked distribution of H/W values, as
seen in figure~\ref{HWg} 
where the concept is illustrated using data from a simulated 
gamma-ray event.

\begin{figure}
\centerline{\includegraphics[width=1.0\textwidth]{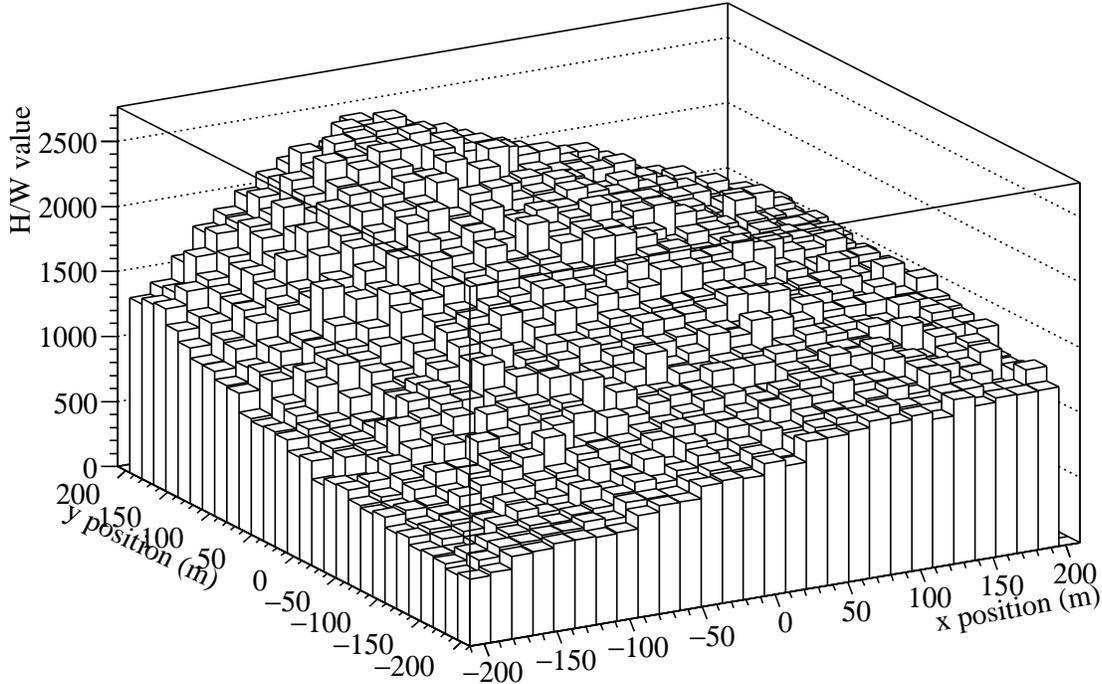}}
\caption{Illustration of the behaviour of the height-to-width ratio of a
single simulated proton 
shower.\label{HWp}}
\end{figure}

Hadronic showers are not expected to exhibit strong peaking, as shown 
in figure~\ref{HWp} 
where the $H/W$ distribution for a typical hadronic shower is shown.
To quantify the flatness of the grid distribution, we calculate $H/W$ for four
separate core locations, each 200 m distant from the best core estimate, along 
orthogonal axes perpendicular to the shower axis.
(200 m was chosen as appropriate to the size of the heliostat array.) 
The average of these four values, $H/W_{200}$, is used to make the ratio
$\xi = \frac{H/W_{200}}{H/W_{max}}$.

\begin{figure}
\centerline{\includegraphics[width=1.0\textwidth]{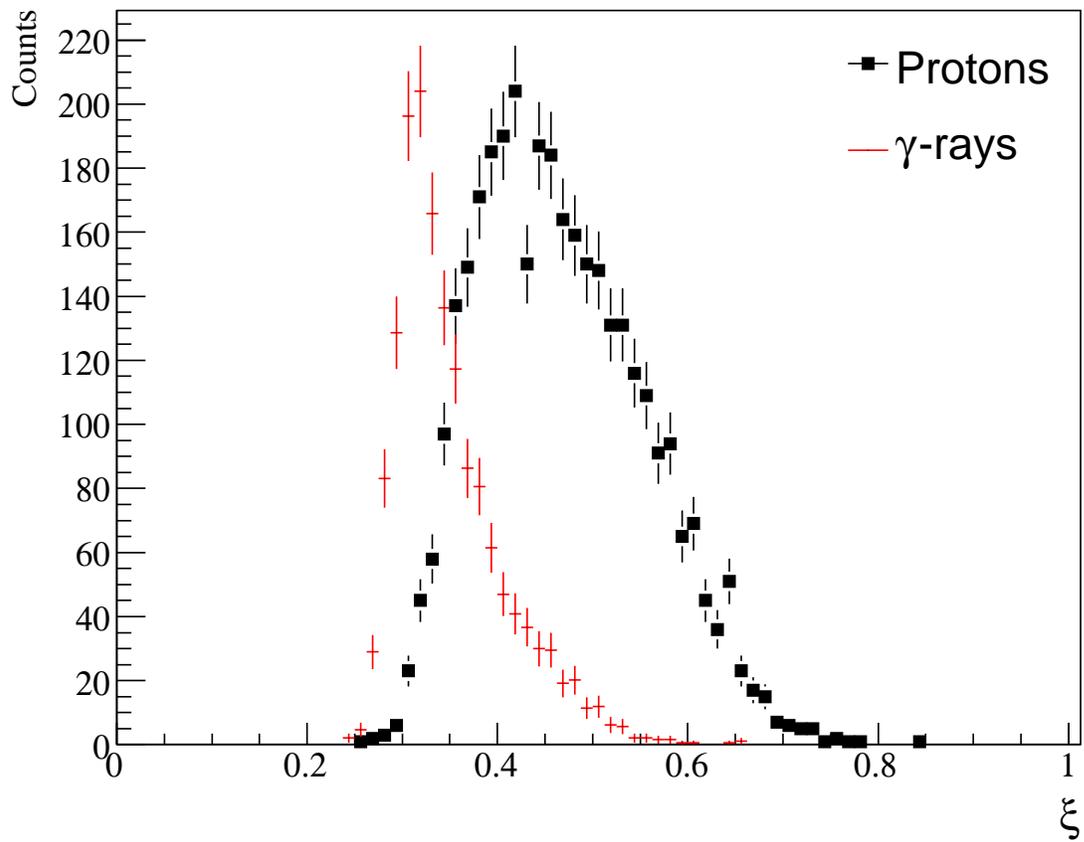}}
\caption{Distribution of the variable $\xi$ (defined in the text)
from simulations of gamma rays and
protons at the Crab transit point.
The curves have been normalized to have the same maximum value.
\label{xi}}
\end{figure}

\begin{figure}
\centerline{\includegraphics[width=1.0\textwidth]{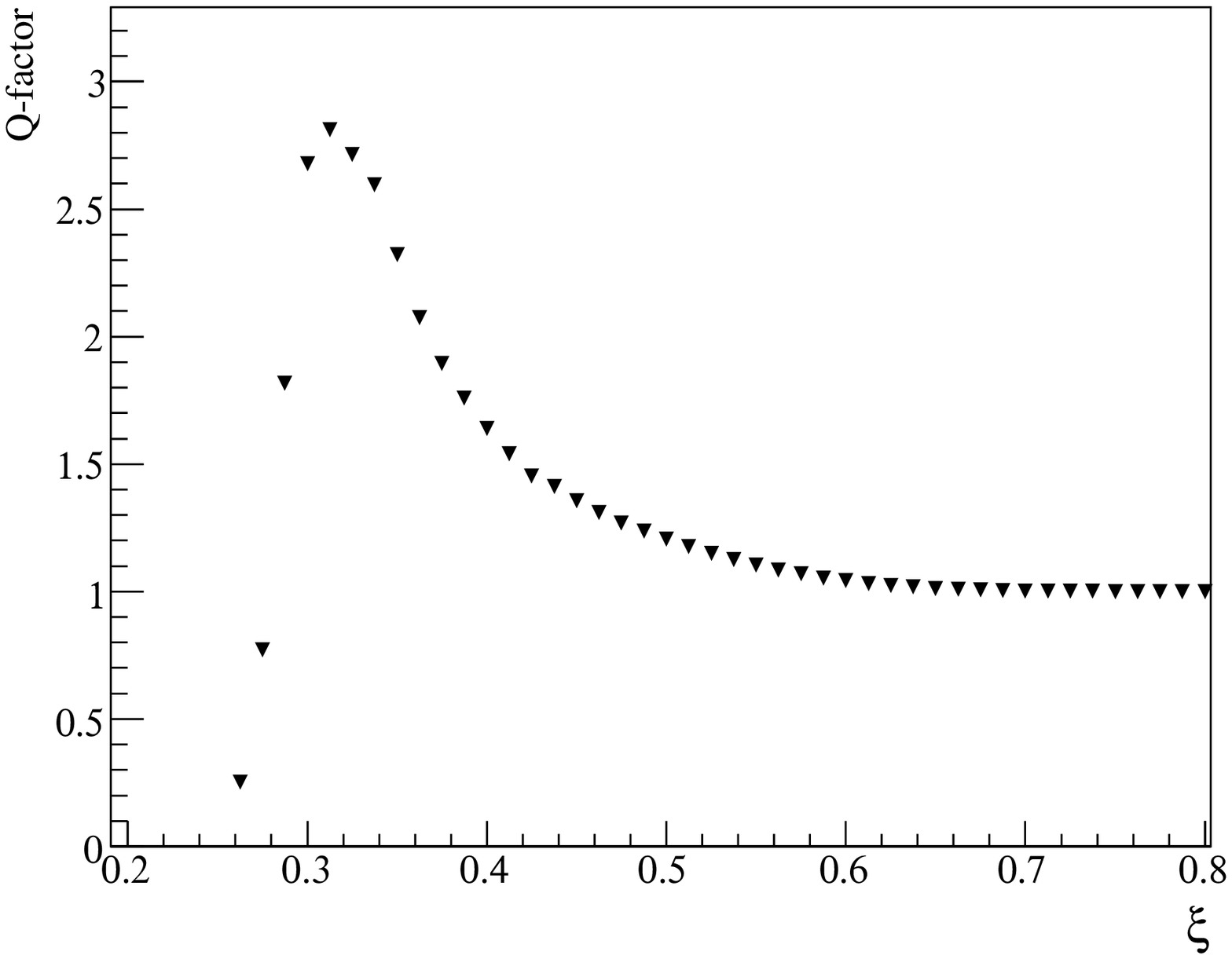}}
\caption{Quality factor $Q$ as a function of $\xi$ cut value, where
$Q = \frac{N'_\gamma/N_\gamma}{\sqrt{N'_h/N_h}}$
with the primed 
quantities being those passing the cut, and the data 
are from simulations 
made at the Crab transit point.
\label{Q}}
\end{figure}

The difference in $\xi$ values for showers due to gamma rays and those 
initiated by protons is evident in figure~\ref{xi} where simulations performed
on a Crab-like spectrum ($E^{-2.4}$) at the Crab transit position are shown.
Gamma-ray showers have, on average, lower values of $\xi$ than do proton
showers. 
Part of this is due to the curvature of the wavefront, as discussed, 
but there is also an effect coming from the thickness of the shower front, 
which is smaller for gamma-ray showers.
The quality factor, $Q$, defined as 
$Q = \frac{N'_\gamma/N_\gamma}{\sqrt{N'_h/N_h}}$
with the primed quantities being those passing a cut on $\xi$, is shown 
as a function of the cut value in figure~\ref{Q}.
A cut value of $\xi < 0.325$ gives the best $Q$ factor (2.6) but only retains
40\% of the gamma rays.
We use a slightly looser cut of $\xi < 0.35$ which keeps 60\% of 
the gamma rays.

As might be expected, there are several biases inherent in the grid ratio
technique.
The variable $\xi$ depends slightly on energy 
over the range explored by STACEE;
its mean value for gamma rays rises from 0.32 at 100 GeV to 0.38 at 1000 GeV.
It also depends on the position of the source on the sky since 
the depth of shower maximum depends on source elevation.
Finally, due to the close packing, and close proximity to the tower, 
of heliostats in the southern part of the field (see figure~\ref{hfield}),
showers with a large fraction of their light hitting those heliostats
have systematically larger values of $\xi$, an effect that must also 
be accounted for.

\begin{figure}
\centerline{\includegraphics[width=1.0\textwidth]{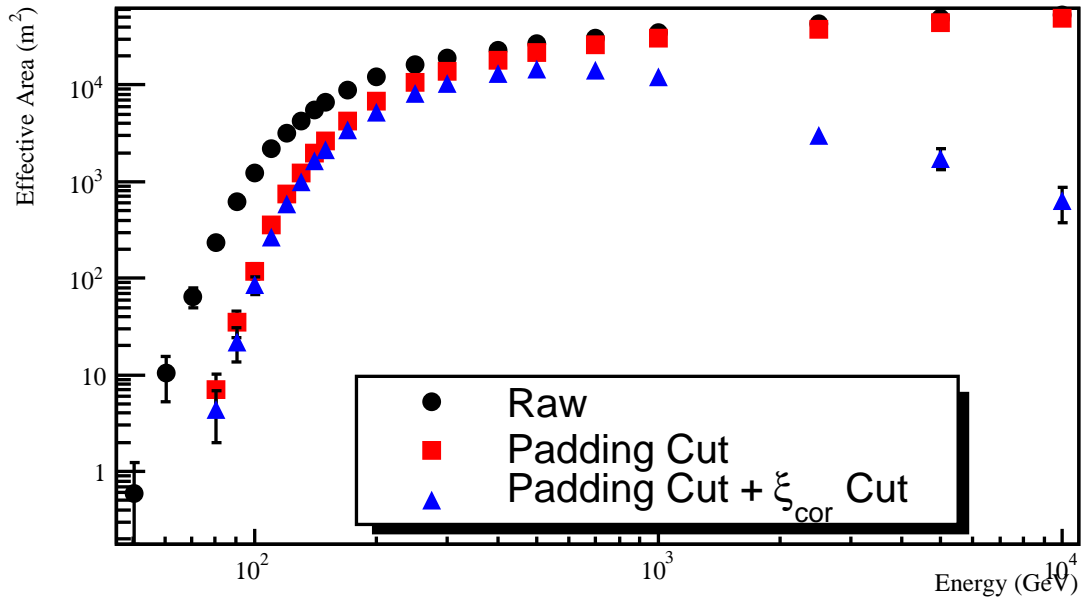}}
\caption{Effective area of the STACEE detector as a function of gamma-ray 
energy with different cuts applied to the data.
Software padding, with its increase in threshold, affects the low energy 
range while hadron suppression lowers the acceptance at high 
energy.
\label{accept}}
\end{figure}

For the purposes of this work, we need to understand these biases 
and their effect on the acceptance of the STACEE detector as a function of
energy and direction. 
We rely on extensive simulations to produce curves like
that shown in figure~\ref{accept} which illustrates, for a representative 
data set (2004 Mrk 421 data), the effective area of STACEE as a function of 
energy with different cuts applied to the data \cite{lindner}. 
Clearly there are two large effects.
At low energies, the increased pulse-height threshold that is part of the
software padding process reduces the acceptance below 100 GeV.
At high energies, the $\xi$ cut lowers the acceptance appreciably above 
1 TeV.
This effect is due to the fact the wavefronts of gamma-ray showers are less 
spherical at higher energies. 

For the two BL Lac objects observed by STACEE, we calculated separate,
source-specific, versions of figure~\ref{accept}.
All these acceptance curves are hour-angle-weighted; an acceptance curve 
was calculated for each of a set of different pointings and these were combined
in an average, weighted according to how long was spent at each pointing.

In summary, the analysis selects data taken under conditions of 
acceptable weather with reliable hardware. 
ON/OFF field brightness differences are removed using library padding and 
hadronic showers are suppressed using the grid ratio technique. 
At this point a signal, manifest as a difference between ON and OFF count rates, is 
sought. 

\section{Results}

\subsection{3C 66A}

The Third Cambridge (3C) radio survey at 159 MHz \cite{cambridge} 
resulted in a 
catalog of 471 sources in the northern hemisphere. 
It was later found that the source numbered 66 was composed of two 
unrelated objects, a BL Lac object now identified as 3C 66A and a radio
galaxy, 3C 66B.
The BL Lac classification has been supported by optical and X-ray 
observations.

3C 66A is coincident with the EGRET source 3EG 0222+4253 \cite{hartman99}, 
but there are other objects in the EGRET error box, including the pulsar
J0218+4232. 
It has been suggested by Kuiper $et~al$ \cite{kuiper} 
that the pulsar contributes to the observed gamma-ray flux 
at energies less than 500 MeV while the BL Lac object dominates
at higher energies.

No confirmed detections of this source have been made at very high energies. 
A result from the Crimean GT-48 telescope \cite{stepanyan} has not 
been confirmed. 
Indeed, measurements in the same energy range have resulted in upper 
limits \cite{whipple66,hegra66},
that are lower than the flux reported by the GT-48 group.
However it is always important to remember that BL Lac objects are highly
variable so a non-confirmation is not necessarily a contradiction.

STACEE observed 3C 66A from September to December, 2003 
as part of a multi-wavelength campaign summarized in Boettcher $et~al$
\cite{boettcher}. 
We acquired a data set of 87 ON/OFF pairs.
Weather and hardware quality cuts removed 31\% of the data,
leaving an ON source live-time of 83.2 ks.
These data have been analysed and presented before \cite{bramel}, but without
applying hadron rejection.

\begin{figure}
\centerline{\includegraphics[width=1.0\textwidth]{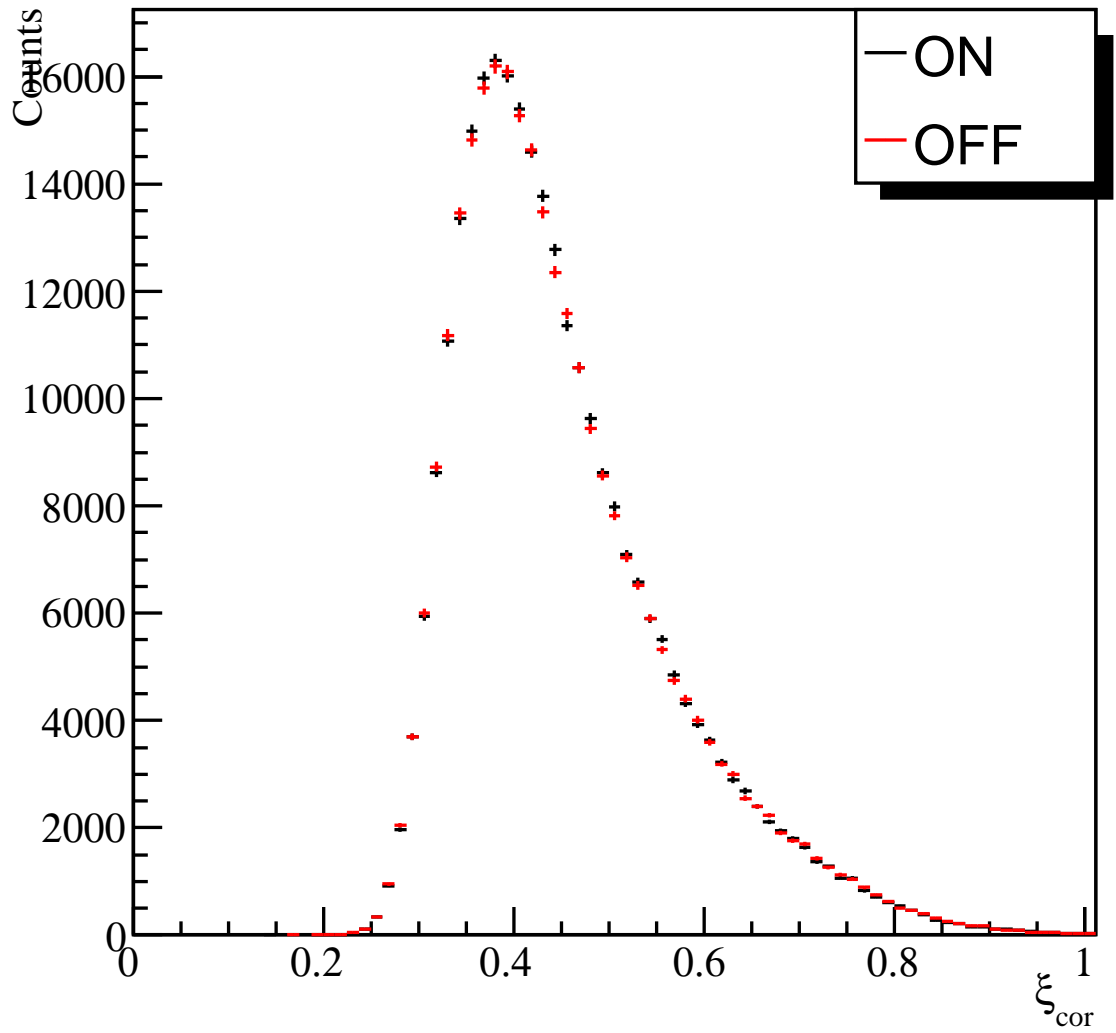}}
\caption{Number of showers, as a function of the grid ratio cut variable $\xi$,
for the 3C 66A data.
The ON and OFF data are essentially overlapping.\label{onoff1}}
\end{figure}

\begin{figure}
\centerline{\includegraphics[width=1.0\textwidth]{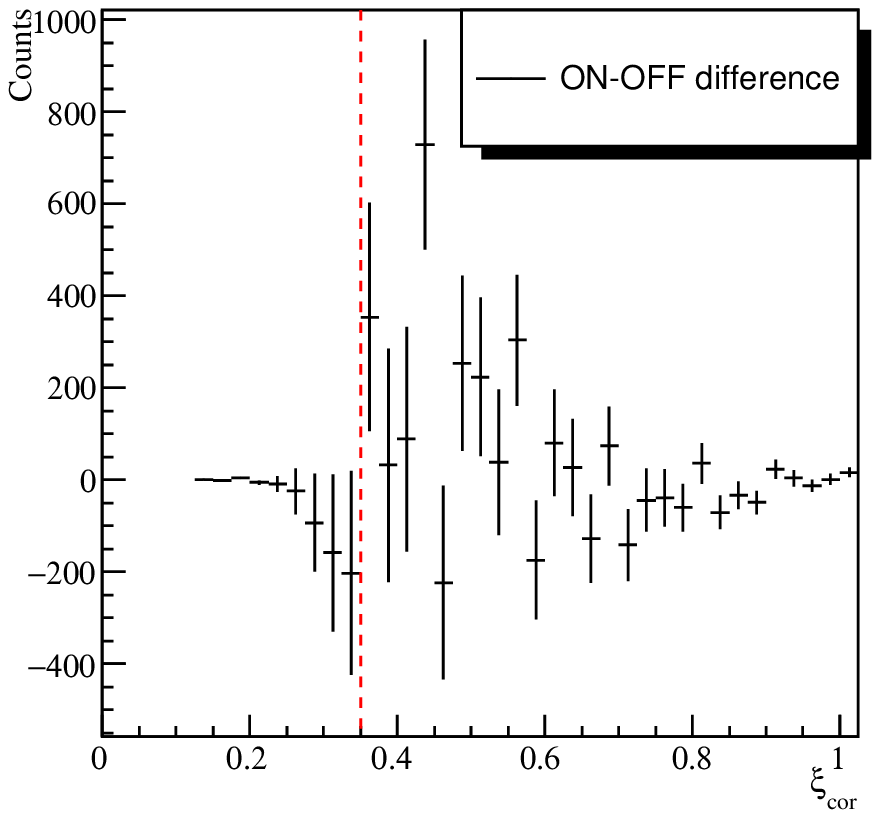}}
\caption{The difference of the two curves plotted in the previous 
figure. The dashed line indicates the value below which gamma-ray events
are expected.\label{onoff2}}
\end{figure}

Figure~\ref{onoff1} shows the $\xi$ distributions for the ON and OFF
data sets and figure~\ref{onoff2} shows the difference. 
There is no evidence for a signal. 
Applying the $\xi < 0.35$ cut results in a net rate of $-0.35 \pm 0.22$
counts per minute, which is 1.6 standard deviations below background.
Since this is consistent with zero we calculate a 
bounded upper limit for the rate.
The $99\%$ CL upper limit on the gamma-ray rate is 0.37 counts per minute.

To calculate an upper limit for the flux, we must make assumptions
about the spectrum of 3C 66A.
Since most VHE sources exhibit power law behaviour over the range of our 
sensitivity, we adopt such a form.
However, it is worth remembering that, depending on the distance of the 
source, EBL attenuation effects can cause a steepening of the spectrum
at high energies and that this can affect our calculated limit.
The choice of spectral index is not well constrained; we choose a value of 
2.5, which is typical for the detected VHE blazars. 
We recognise that our result is tightly correlated to this value.

\begin{figure}
\centerline{\includegraphics[width=1.0\textwidth]{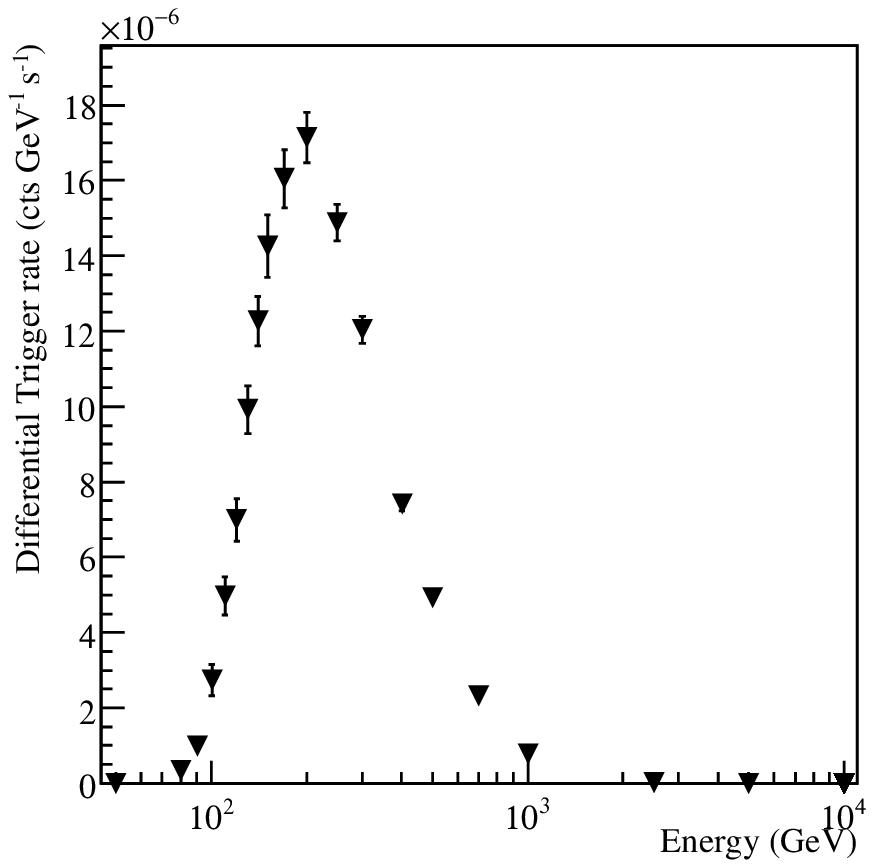}}
\caption{The convolution of power law energy spectrum (spectral 
index = 2.5) with
the hour-angle weighted effective area curve for the 3C 66A data set.
The peak at 185 GeV provides an operational definition of the energy 
threshold. \label{response}}
\end{figure}

Assuming the spectral form, $dN/dE = N_0(E/E_0)^{-2.5}$ 
and convoluting it with the effective area $vs$ energy curve results in the
response curve shown in figure~\ref{response}.
It is customary in VHE gamma-ray astronomy to define the peak value of this 
curve to be the energy threshold. 
According to this definition, the energy threshold for this measurement is 
$185$ GeV, with a systematic uncertainty of $45$ GeV.

\begin{figure}
\centerline{\includegraphics[width=1.0\textwidth]{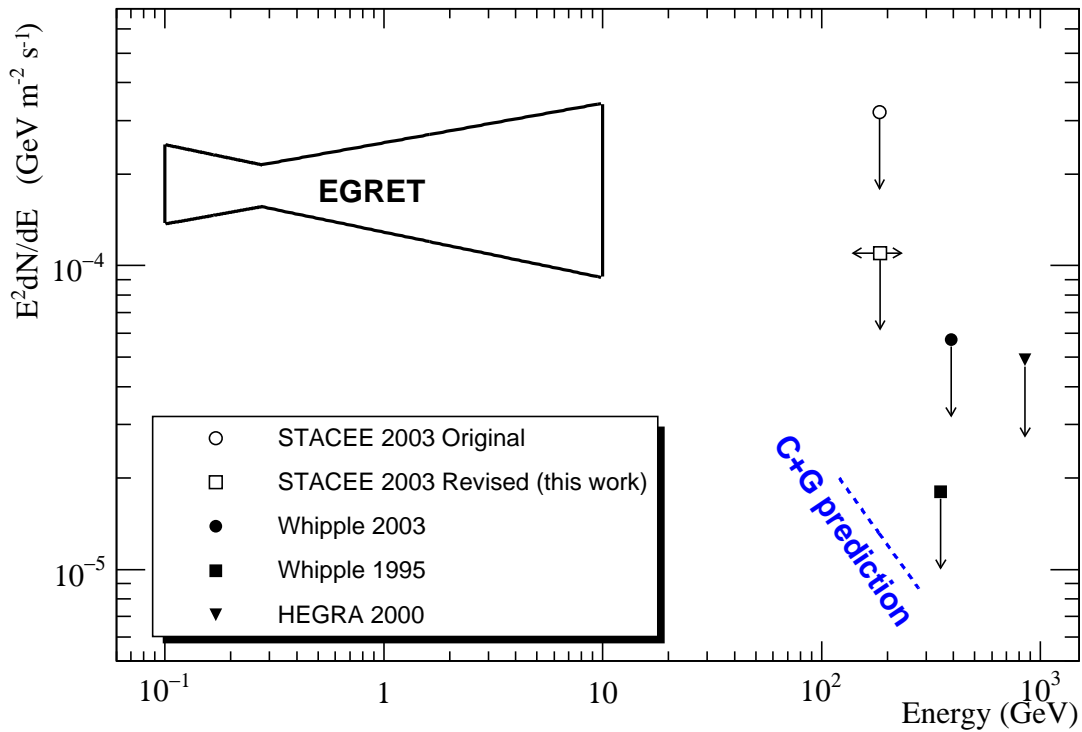}}
\caption{The spectral energy distribution for 3C 66A in the gamma-ray region.
Note that all small symbols denote upper limits.
The STACEE limit (revised) has been improved using the technique described 
in this paper.   
The EGRET symbol corresponds to data from 3EG 0222+4253 which includes 3C 66A
but also a pulsar. \label{sed1}}
\end{figure}

Our 99\% CL upper limit on the gamma-ray flux at 185 GeV is 
$E^2 dN/dE$ (185 $\pm 45_{sys}$ GeV) $< 1.1 \times 10^{-4}$ GeV m$^{-2}$s$^{-1}$.
Stated differently, we can say that the 3C 66A integral gamma-ray 
flux above 185 GeV 
is less than 15\% of that of the Crab Nebula, using results from STACEE 
measurements of the Crab over the same energy range (\cite{fortin}).
As can be seen in figure~\ref{sed1}, this upper limit is approximately 
three times lower than our previous result
\cite{bramel} due to the improved hadron rejection afforded by the grid ratio
technique. 
However, it is still well above the flux value obtained by interpolating 
the predictions found in the study by Costamante and Ghisellini 
\cite{Costamante}.
Our point is above those from the imaging Cherenkov telescopes, Whipple and
HEGRA, but it is at a lower energy, where the EBL absorption of photons from 
this distant source is expected to be less important.

The EGRET measurement shown in the figure is actually from the source 
3EG 0222+4253 which includes the pulsar J0218+4232 \cite{kuiper} 
as well as 3C 66A. 
It is not clear how much of the observed flux can be attributed to 3C 66A
alone.

\subsection{OJ 287}

OJ 287 was discovered in 1968 in the Ohio State University survey of radio 
sources at 1415 MHz \cite{dixon}. 
It was soon identified with an optical source of magnitude 14.5 \cite{blake}
and further work at other wavelengths established its membership in the BL Lac
class \cite{strittmatter}. 
Its redshift has been measured to be 0.306 \cite{miller,sitko}.

Optically, OJ 287 has been followed for more than 100 years. It appears to have 
a marked increase in optical emission every twelve years, 
a phenomenon that can
be explained by supposing that the AGN contains a pair of orbiting black holes,
a primary with mass $10^{10}~M_{\bigodot}$ and a secondary with mass 
$10^7~M_{\bigodot}$ \cite{sillanpaa}.
The last outburst occured in 1994-95, and it was during this time that EGRET 
accumulated most of its observing time on the source.
EGRET detected OJ 287 as a relatively weak source with an average integral 
flux of $10.6 \pm 3.0 \times 10^{-8}$ photons cm$^{-2}$ s$^{-1}$ 
\cite{hartman99}.
This source has not been detected by any of the VHE telescopes.

STACEE observed OJ 287 from December, 2003 to February, 2004 and obtained a 
data set of 28 ON/OFF pairs. 
After weather and other quality cuts were applied, 52\% of the data remained,
corresponding to 21.1 ks of ON-source live time.
The distributions of $\xi$ for ON and OFF are very similar to those shown in 
figure~\ref{onoff1} and the difference distribution resembles 
figure~\ref{onoff2}. 
The standard $\xi$ cut results in a flux of 0.35 $\pm$ 0.39 counts min$^{-1}$
and a statistical significance of 0.9 standard deviations above background. 
As with 3C 66A, we use these numbers to calculate a 99\% CL upper limit on the 
gamma-ray rate. 
This upper limit is 1.29 photons per minute.
Using the hour-angle-weighted effective area curve and assuming a power law
spectrum with index of 2.5, we obtain an energy threshold of 145 $\pm$ 36
GeV and a gamma-ray flux upper limit of
$E^2 dN/dE$ (145 $\pm 36_{sys}$ GeV) $< 4.0 \times 10^{-4}$ GeV m$^{-2}$s$^{-1}$.
This corresponds to 52\% of the Crab Nebula flux above the same energy.

\begin{figure}
\centerline{\includegraphics[width=1.0\textwidth]{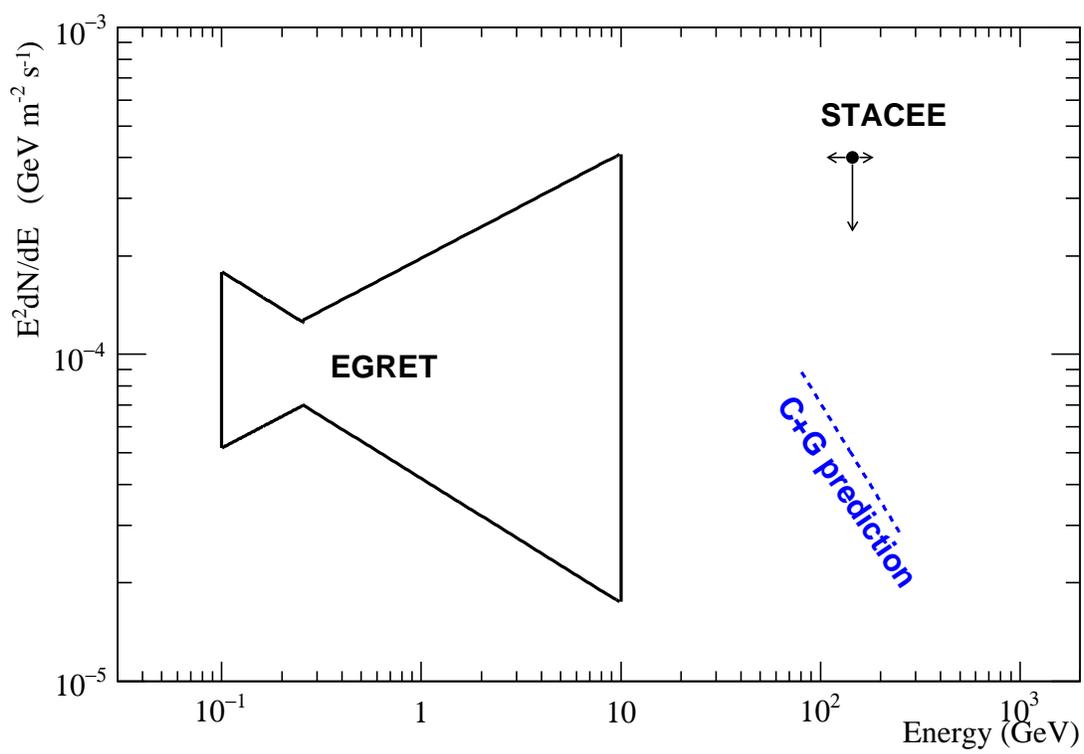}}
\caption{The spectral energy distribution for OJ 287 in the gamma-ray region.} 
\label{sed2}
\end{figure}

The spectral energy distribution, as measured by EGRET and with the STACEE 
limit and Costamante and Ghisellini prediction included, is shown 
in figure~\ref{sed2}.
It is seen that our measurements are above the Costamante and Ghisellini 
prediction but are the only ones reported at these energies.

\section{Conclusions}

We have presented data from recent STACEE observations of two of the three
LBL candidates suggested as potential TeV emitters by Costamante and 
Ghisellini \cite{Costamante}.
We have not detected a signal from either of these sources and have set upper
limits on their gamma-ray flux levels.
Although the sensitivity of STACEE has been improved through the use of a new
hadron rejection technique, also presented in this paper, we cannot rule
out emission at the level suggested in \cite{Costamante}.
Although neither of the sources has yet been detected in the VHE range, 
the third LBL candidate, BL Lacertae has recently been detected by the 
MAGIC collaboration \cite{albert07}.
Their results indicate that BL Lacertae is highly variable, as is typical for 
most blazars.
We look forward to improved limits or detections now that a new generation of 
ground-based detectors, VERITAS \cite{holder} and MAGIC \cite{lorenz}, are viewing the 
northern sky with improved sensitivity and lower energy thresholds.

\section{Acknowledgements}

We are grateful to the staff at the National Thermal Solar Test Facility 
for their enthusiastic and professional support. 
This work was funded in part by the U.S. National Science Foundation, 
the University of California, Los Angeles,
the Natural 
Sciences and Engineering Research Council, le Fonds Qu\'ebecois de la 
Recherche sur la Nature et les Technologies,
the Research Corporation, and the 
California Space Institute.

\end{document}